\documentstyle[preprint,eqsecnum,aps,version2]{revtex}

\def\331{\rm SU(3)_L \times U(1)_X}
\def\sm{\rm SU(2)_L \times U(1)_Y}
\def\Q{\rm U(1)_Q}
\def\noalignand{
             \noalign{\vbox{\vskip\abovedisplayskip \hbox{and}
              \vskip\belowdisplayskip}}}
\def\eff{{\it eff}}
\def\spp{\sin\alpha_{++}}
\def\cpp{\cos\alpha_{++}}
\def\tpp{\tan\alpha_{++}}
\def\sspp{\sin^2\alpha_{++}}
\def\ccpp{\cos^2\alpha_{++}}
\def\sp{\sin\alpha_+}
\def\cp{\cos\alpha_+}
\def\tp{\tan\alpha_+}
\def\ssp{\sin^2\alpha_+}
\def\ccp{\cos^2\alpha_+}
\def\Re{\,{\rm Re}\,}
\def\Im{\,{\rm Im}\,}
\def\Mp{M_{Y^+}}
\def\Mpp{M_{Y^{++}}}
\def\mpp{m_{H^{++}}}
\def\mp{m_{H^+}}
\begin{document}
\draft
\preprint{IFP-460-UNC}
\preprint{TRI-PP-93-11}
\preprint{hep-ph/9302271}
\preprint{February 1993}
\preprint{revised, June 1993}
\vspace{-0.5cm}
\begin{title}
\begin{center}
$Z$--$Z'$ mixing and oblique corrections \\
in an ${\rm SU(3) \times U(1)}$ model
\end{center}
\end{title}

\author{James T. Liu}
\begin{instit}
\begin{center}
Institute of Field Physics, Department of Physics and Astronomy,\\
University of North Carolina, Chapel Hill, NC  27599--3255, USA
\end{center}
\end{instit}
\author{Daniel Ng}
\begin{instit}
\begin{center}
TRIUMF, 4004 Wesbrook Mall\\
Vancouver, B.C., V6T 2A3, Canada
\end{center}
\end{instit}
%
%
\begin{abstract}
{\baselineskip=20pt
We address the effects of the new physics predicted by the $\331$ model on
the precision electroweak measurements.  We consider
both $Z$--$Z'$ mixing and one-loop oblique corrections, using a combination
of neutral gauge boson mixing parameters and the parameters $S$ and $T$.
At tree level, we obtain strong limits on the $Z$--$Z'$ mixing angle,
$-0.0006<\theta<0.0042$ and find $M_{Z_2}>490 {\rm GeV}$ (both at 90\% C.L.).
The radiative corrections
lead to $T>0$ if the new Higgs are heavy, which bounds the Higgs masses to be
less than a few TeV.  $S$ can have either sign depending on the Higgs
mass spectrum.  Future experiments may soon place strong restrictions on this
model, thus making it eminently testable.

}
\end{abstract}
\newpage

\section{Introduction}
Recently a model based on the gauge group $\331$ has been proposed as a
possible explanation of the family replication question \cite{frampton}.
By matching the gauge coupling constants at the electroweak scale
\cite{ng}, the mass of the new
heavy neutral gauge boson, $Z'$, is bounded to be less than $2.2$ TeV and
the mass upper bound for the new charged gauge bosons, $Y^{\pm\pm}$ and
$Y^{\pm}$, is $435$ GeV \cite{massb}.  Since $Y^{++}$ and $Y^{+}$ carry two
units of
lepton number, they are called dileptons.  Unlike most extensions of the
standard model, in which the masses of the new gauge bosons are not bounded
from above, this model would be either realized or ruled out in the future
high energy colliders such as the superconducting supercollider and the next
linear collider.

The new $Z'$, by mixing with the standard model neutral gauge boson $Z$,
modifies the neutral current parameters as well as the $\rho$-parameter
\cite{langacker}.  The dileptons, $Y^{\pm\pm}$ and $Y^{\pm}$, and the
new charged Higgs, $H^{\pm\pm}$ and $H^{\pm}$, on the other hand, do not
participate directly in the precision LEP experiments \cite{lep} nor the
neutrino scattering experiments \cite{charm}.  Instead, they only enter
radiatively, mainly via their oblique corrections to the $W^{\pm}$ and $Z$
propagators \cite{peskin,altarelli,marciano,kennedy,holdom,golden}.
Nevertheless, such radiative corrections may be comparable to the tree level
corrections due to the $Z$--$Z'$ mixing.  Thus we treat both cases in the
following.

If the masses of the dileptons are degenerate, we may expect the oblique
corrections
to vanish.  However, the mass degeneracy is lifted when $\sm$ breaks
into $\Q$; thus the mass squared splitting would be on the order of
$M_W^2$.  As a result, the oblique corrections to the parameters $S$ and $T$
\cite{peskin}
are expected to be on the order of $(1/\pi)(M_W^2 / M_{Y^{++}}^2)$, where
$M_{Y^{++}}$ is the mass of $Y^{\pm\pm}$.  In addition, oblique
corrections due to the new heavy charged Higgs, $H^{\pm\pm}$ and $H^{\pm}$,
are induced by the small mixing between Higgs multiplets.  The contributions
have the general form
$(1/\pi)(M_W^2 / M_{Y^{++}}^2)(m_H^2/M_{Y^{++}}^2)$, where $m_H$ is the mass
of the new charged Higgs.  Hence, the heavy charged Higgs contributions
would be important even when the dilepton mass splitting is small.

Our analysis in this paper concentrates on both tree level and one-loop
oblique corrections to the standard model due to the new physics of the
$\331$ model.  For the dileptons and the new Higgs, which only
contribute radiatively, we use the $S$, $T$ and $U$ parameters.  However the
effects of the $Z'$, which enters at tree level, cannot be fully
incorporated into this formalism, and may instead be parametrized by a
$Z$--$Z'$ mixing angle as well as the mass of the heavy $Z_2$.  We thus use
five parameters to describe the new physics: the two $Z'$ parameters and the
three oblique ones.  Starting
with a discussion of tree level mixing, we perform a five
parameter fit to experimental data to put strong limits on the $Z$--$Z'$
mixing angle.  We then discuss the consequences of the fit on the other
particles by carrying out a complete
one-loop calculation of $S$ and $T$ for dilepton gauge bosons
and the new Higgs bosons.  The new quarks, which are SU(2) singlets,
do not contribute.

\section{Tree level mixing}
We first outline the model, following the notation given in
\cite{ng}. The fermions transform under $\rm SU(3)_c\times\331$
according to
\def\bold#1{{\bf #1}}
\begin{mathletters}
\begin{eqnarray}
\psi_{1,2,3} = \pmatrix {e\cr \nu_e\cr e^c\cr} \: , \pmatrix {\mu \cr
\nu_\mu
   \cr \mu^c \cr} \: , \pmatrix {\tau \cr \nu_\tau \cr \tau^c \cr}
     \qquad &\mathbin:& \qquad (\bold1, \: \bold3^\ast, \: 0) \ , \\
Q_{1,2} = \pmatrix {u\cr d\cr D\cr} \: , \pmatrix {c\cr s\cr S\cr}
     \qquad &\mathbin:&\qquad(\bold3,\: \bold3,\:
	-\textstyle\frac{1}{3})\ ,\\
Q_3 = \pmatrix {b\cr t\cr T\cr} \qquad &\mathbin:& \qquad
	(\bold3, \: \bold3^\ast,\: \textstyle\frac{2}{3} ) \ ,  \\
d^c, \: s^c, \: b^c \qquad &\mathbin:& \qquad (\bold3^\ast,\:\bold1,\:
	\textstyle\frac{1}{3}) \ ,  \\
u^c, \: c^c, \: t^c \qquad &\mathbin:& \qquad(\bold3^\ast,\:\bold1,\:
	-\textstyle\frac{2}{3}) \ , \\
D^c, \: S^c \qquad &\mathbin:&  \qquad (\bold3^\ast,\:\bold1,\:
	\textstyle\frac{4}{3}) \ ,  \\
T^c \qquad&\mathbin:&\qquad(\bold3^\ast,\:\bold1,\:\textstyle-\frac{5}{3})\ .
\end{eqnarray}
\end{mathletters}
where $D$, $S$ and $T$ are new quarks with charges $-4/3$, $-4/3$ and $5/3$
respectively.  The minimal Higgs multiplets required for the
symmetry breaking hierarchy and fermion masses are given by
\begin{mathletters}
\begin{eqnarray}
\label{eq:Hmult}
\Phi = \pmatrix{\phi^{++}\cr \phi^+\cr \phi^0\cr} \qquad \qquad
         &\mathbin:& \qquad (\bold1,\,\bold3,\,1) \ ,  \\
\Delta = \pmatrix{\Delta_1^+\cr \Delta^0\cr \Delta_2^-\cr} \qquad \qquad
         &\mathbin:& \qquad (\bold1,\,\bold3,\,0) \ , \\
\Delta ' = \pmatrix{\Delta '^0\cr \Delta '^-\cr \Delta '^{--}\cr} \qquad
       \quad &\mathbin:& \qquad (\bold1,\,\bold3,\,-1) \ ,  \\
\noalignand
\eta = \pmatrix {\eta_1^{++} & \eta_1^+/\sqrt{2} & \eta^0/\sqrt{2} \cr
                 \eta_1^+/\sqrt{2} & \eta_1 ^0 & \eta^-/\sqrt{2} \cr
                 \eta^0/\sqrt{2} & \eta^-/\sqrt{2} & \eta^{--} \cr}
      \qquad &\mathbin:& \qquad (\bold1,\,\bold6,\,0)
\ .
\end{eqnarray}
\end{mathletters}

The non-zero vacuum expectation value (VEV) of $\phi^0$, $u/\sqrt{2}$, breaks
$\331$ into $\sm$.  The SU(2) components of $\Delta$ and $\Delta'$ then
behave like the ordinary Higgs doublets of a two-Higgs standard model.
The sextet, $\eta$,
is required to obtain a realistic lepton mass spectrum.  For simplicity,
we will assume its VEVs are zero.  As $\331$ is
broken into $\sm$, the sextet will decompose into an SU(2) triplet, an
SU(2) doublet and a charged SU(2) singlet.  We will also assume that the mass
splitting of these scalars within their multiplets is small; hence their
contributions to $S$ and $T$ will be negligible.

As $\sm$ is broken by the VEVs of $\Delta^0$ and $\Delta'^0$,
$v/\sqrt{2}$ and $v'/\sqrt{2}$, they will provide masses for
the standard model gauge bosons, $W^\pm$ and $Z$.  The VEVs also induce
$Z$--$Z'$ mixing as well as
the mass splitting of $Y^{\pm\pm}$ and $Y^{\pm}$.  Hence we obtain the
masses for the charged gauge bosons,
\begin{mathletters}
\begin{eqnarray}
M^2_W &=& \textstyle\frac{1}{4} \: g^2 (v^2 + v'^2) \ , \\
M^2_{Y^+} &=& \textstyle\frac{1}{4} \: g^2(u^2 + v^2) \ ,
\\
\noalignand
M^2_{Y^{++}} &=& \textstyle\frac{1}{4} \: g^2 (u^2 + v'^2 ) \ ,\phantom{a}
\end{eqnarray}
\end{mathletters}
and the mass-squared matrix for $\{Z, \:  Z'\}$
\begin{eqnarray}
{\cal M}^2 = \pmatrix {M^2_Z & M^2_{ZZ'}\cr
M^2_{ZZ'} & M^2_{Z'}}\ ,
\end{eqnarray}
with
\begin{mathletters}
\begin{eqnarray}
M^2_Z &=& \frac{1}{4} \frac{g^2}{\cos^2\theta_W} \: (v^2+v'^2) \ ,  \\
M^2_{Z'} &=& \frac{1}{3} g^2 \left[
              \frac{\cos^2\theta_W}{1-4 \sin^2\theta_W} \: u^2
           \: + \:  \frac{1-4 \sin^2\theta_W}{4\cos^2\theta_W}\: v^2
  \hfill \right. \nonumber \\
          &&\qquad\left.
          \: +\: \frac{(1+2\sin^2\theta_W)^2}
                   {4 \cos^2\theta_W(1-4 \sin^2\theta_W)}\: v'^2
         \right] \ ,  \\
M^2_{ZZ'} &=& \frac{1}{4\sqrt{3}} \, g^2 \left[
         \frac{\sqrt{1-4 \sin^2\theta_W}}{\cos^2\theta_W}\: v^2 -
        \frac{1+2\sin^2\theta_W}
            {\cos^2\theta_W \sqrt{1-4\sin^2\theta_W}}\: v'^2 \right]  .
\end{eqnarray}
\end{mathletters}
The mass eigenstates are
\begin{mathletters}
\begin{eqnarray}
Z_1 &=& \cos\theta\ Z - \sin\theta\ Z' \ ,  \\
\noalignand
Z_2 &=& \sin\theta\ Z + \cos\theta\ Z' \ ,
\end{eqnarray}
\end{mathletters}
where the mixing angle is given by
\begin{equation}
\label{angle}
\tan^2\theta=\frac{M_Z^2-M_{Z_1}^2}{M_{Z_2}^2-M_Z^2} \ .
\end{equation}
with $M_{Z_1}$ and $M_{Z_2}$ being the masses for $Z_1$ and $Z_2$.
Here, $Z_1$ corresponds to the standard model neutral gauge boson and
$Z_2$ corresponds to the additional neutral gauge boson.
For small mixing, we find $\theta \approx {M^2_{ZZ'}}/{M^2_{Z_2}}\ll1$.

Since $M_{Z_1}$ has been precisely determined by the LEP experiments, the
new contributions are parametrized by the two $Z'$ parameters, $M_{Z_2}$ and
$\theta$.  The structure of the minimal Higgs sector gives additional
constraints on the allowed region of $(M_{Z_2},\theta)$ parameter space,
and forces $\theta\ll1$ for $M_{Z_2}\gg M_{Z_1}$.  However, we will not make
use of this constraint so as to allow for extended Higgs sectors.

While we have only been discussing tree level relations so far, it is
important to include both the standard model and new
radiative corrections as well.  We take the oblique
corrections into account by using the starred functions of Kennedy and Lynn
\cite{kennedy2}.  Following \cite{peskin}, the effect of new heavy particles
on the starred functions may be expressed in terms of $S$, $T$ and $U$.
The effects of the tree level $Z$--$Z'$ mixing and the presence of the
new $Z_2$ gauge boson can then be expressed as shifts of the starred
functions.  We ignore effects due to the combination of both
mixing and radiative corrections, as they are suppressed.

In order to perform a fit to experiment, we need to express the $\331$ model
predictions in terms of both tree level $Z'$ parameters, $(M_{Z_2},\theta)$,
and one-loop parameters, $(S,T,U)$.
This is most easily done by first calculating the
standard model observables with the addition of $S$, $T$ and $U$ and
then shifting the
results by the tree level parameters.  We consider both (i) $Z$-pole
experiments which are sensitive to the mixing only
and (ii) low energy experiments which are sensitive to both mixing and the
presence of the $Z_2$.  The experimental values that we use for the five
parameter fit, along with the standard model predictions
(for reference values of $m_t=150\rm GeV$ and $m_H=1000\rm GeV$
\cite{peskin}),
are given in table \ref{data}.  For the $Z$-pole data, $M_W/M_Z$ and
$Q_W(Cs)$, we use the values given in Ref.~\cite{langacker2}, while $g_L^2$
and $g_R^2$ are given in Ref.~\cite{pdb}.
We find it convenient to approximate the top quark and standard
model Higgs mass dependence through shifts in $S$, $T$ and $U$.

The new contributions to the measurable quantities due to the presence
of the $Z_2$ and $Z$--$Z'$ mixing are given in the appendix.  For the
$(S,T,U)$ dependence of the observables, we use the results given in
Ref.~\cite{peskin}.  The result of the fit in the $(M_{Z_2},\theta)$ plane
(with $S$, $T$ and $U$ unrestricted) is presented in Fig.~\ref{fig1} and
indicates
that $Z$--$Z'$ mixing is highly restricted.  This is partially due to the
large couplings of the $Z'$ to quarks.  At 90\% C.L., we find
$-0.0006<\theta<0.0042$ and $M_{Z_2}>490\rm GeV$.  Note the latter
restriction is comparable to that obtained from tree level FCNC
considerations in the quark sector.

Although not used in the fit, the minimal Higgs sector leads to further
restrictions on the $Z_2$ mass and mixing.  The constraint on the $Z$--$Z'$
mixing is shown by the dotted line in Fig.~\ref{fig1}.  Due to the symmetry
breaking hierarchy, $u\gg v,v'$, the dilepton and $Z_2$ masses are related.
Using the limit $M_{Y^+}>300\rm GeV$ from polarized muon
decay \cite{carlson,beltrami},
we find $M_{Z_2}>1.4\rm TeV$, as indicated on the figure.  Because of the
upper bound on $\331$ unification,
$M_{Z_2}$ must be below $2.2\rm TeV$, thus
giving a narrow window for the allowed $Z_2$ mass.

The presence of the $Z_2$ gauge boson affects the fit in the $S$--$T$
plane as shown in Fig.~\ref{fig2}.  We see that the tree level mixing may
appear as effective contributions to $S$ and $T$.  The dominant effect
is to give a positive contribution to $T$ due to the downshift in the
$Z_1$ mass.  The large region of negative $T$ corresponds to high $Z_2$
mass and small mixing.  Imposing an upper bound on $M_{Z_2}$ will affect the
fit in this region.  At 90\% C.L. we find
\begin{equation}
-1.34~\leq S~\leq~0.28\ , \qquad -3.07~\leq T~\leq~0.45\ ,
\label{STlimit}
\end{equation}
keeping in mind that the errors are nongaussian.
Although the definitions of $S$, $T$ and $U$ are model independent,
these numbers are
valid only for the $\331$ model due to the tree level effects.  We use these
results in the next section to constrain the new charged Higgs masses.

\section{Radiative corrections}
The radiative corrections arising from the dileptons and the new heavy
Higgs are process independent and may be parametrized by $S$, $T$ and $U$.
Following the notation of \cite{peskin}, we define
%
\begin{mathletters}
\begin{eqnarray}
S&=&16\pi\left[\Pi_{33}'(0)-\Pi_{3Q}'(0)\right]\ ,\\
T&=&{4\pi\over \sin^2\theta_W M_W^2}\left[\Pi_{11}(0)-\Pi_{33}(0)\right]\ ,\\
U&=&16\pi\left[\Pi_{11}'(0)-\Pi_{33}'(0)\right]\ .
\end{eqnarray}
\end{mathletters}
In the above, the vacuum polarizations, $\Pi(q^2)$, and their derivatives
with respect to $q^2$, $\Pi'(q^2)$, include only new physics beyond the
standard model.  Implicit in this parametrization is the assumption that
the scale of new physics is much larger than $M_Z$.


The $\331$ model predicts three classes of new particles: the new quarks
$D$, $S$ and $T$, new gauge bosons, $Y^{\pm\pm}$, $Y^\pm$ and $Z'$ and new
Higgs scalars.  Since the new quarks are SU(2) singlets, they do not enter
into the oblique corrections which are only sensitive to SU(2) electroweak
physics.  Similarly, $Z'$ will not contribute except through $Z$--$Z'$
mixing which was addressed in the previous section.  Thus in the limit of
small mixing, only dileptons and new Higgs particles will contribute
radiatively to $S$ and $T$ (in addition to the deviations of the top
quark and standard model Higgs masses from their reference values).
Because of spontaneous symmetry
breaking, we must examine the new gauge and Higgs sector simultaneously.

In order to simplify the analysis of the Higgs sector, we assume that
the sextet $\eta$ does not acquire a VEV.  As a result it can be treated
separately from the dileptons, and we now focus on the three SU(3) triplet
Higgs, (\ref{eq:Hmult}--c).  These three Higgs contain a total of 18 states
of which 8 are ``eaten up'' by the Higgs mechanism to give masses to the
various gauge bosons.  Ignoring $Z$--$Z'$ mixing, the SU(2) doublets coming
from $\Delta$ and $\Delta'$ form a standard two-Higgs model with
$\tan\beta=v'/v$ and five physical Higgs particles, $h^\pm$, $a^0$ and
$h^0_{1,2}$ \cite{Hunt}.  The remaining 5 Higgs are given by
\begin{mathletters}
\begin{eqnarray}
\label{eq:higgs}
H^{\pm\pm}&=&\spp\phi^{\pm\pm}+\cpp\Delta^{\prime\pm\pm}\ ,\\
H^\pm&=&\sp\phi^++\cp\Delta_2^+\ ,\\
H^0&=&\sqrt{2}\Re\phi^0\ ,
\end{eqnarray}
\end{mathletters}
where we have defined the ratio of VEVs as $\tpp=v'/u$ and $\tp=v/u$.
These two VEV angles and $\tan\beta$ are not independent, but are related
by $\tan\beta=\tpp/\tp$.
Orthogonal to these states are the would be Goldstone bosons
\begin{mathletters}
\begin{eqnarray}
\pi^{\pm\pm}&=&\cpp\phi^{\pm\pm}-\spp\Delta^{\prime\pm\pm}\ ,\\
\pi^\pm&=&\cp\phi^+-\sp\Delta_2^+\ ,\\
\pi^0&=&\sqrt{2}\Im\phi^0\ ,
\end{eqnarray}
\end{mathletters}
corresponding to $Y^{\pm\pm}$, $Y^\pm$ and $Z'$ respectively.
Again we have assumed the $Z$--$Z'$ mixing is not important for one-loop
oblique corrections.

Since the two-Higgs model has already been considered in detail (see for
example Ref.~\cite{2higgs,2higgs2}), we will only
focus on the dileptons and additional Higgs.  Assuming the
symmetry breaking hierarchy $u\gg \{v,v'\}$, we see that $\{\tpp,\tp\}\ll 1$
so that $H^{\pm\pm}$ and $H^\pm$ are mostly SU(2) singlets, and the
would be Goldstone bosons giving masses to the dilepton doublet
($Y^{++}$, $Y^+$)
are mostly contained in the $\Phi$ doublet ($\phi^{++}$, $\phi^+$).
Although the mixings between the SU(2) singlet and doublet scalars are
small, the oblique corrections can be important as their contributions
are not protected by the custodial symmetry.

Let us first consider only the contributions from the dilepton gauge bosons
$(Y^{++},Y^+)$ which corresponds to the limit
$\{\tpp,\tp\}\to0$.  In this limit, the new Higgs, (\ref{eq:higgs}--c), are all
SU(2) singlets and only the dilepton doublet contributes to $S$, $T$ and $U$.
We find
\begin{mathletters}
\begin{eqnarray}
\label{eq:limSTU}
S&=&-{9\over4\pi}\ln{\Mp^2\over\Mpp^2}\ ,\\
T&=&{3\over16\pi\sin^2\theta_WM_W^2}F(\Mp^2,\Mpp^2)\ ,\\
U&=&-{1\over4\pi}\biggl[-{19\Mp^4-26\Mp^2\Mpp^2+19\Mpp^4
	\over3(\Mp^2-\Mpp^2)^2}	\nonumber\\
&&\qquad\qquad+{3\Mp^6-\Mp^4\Mpp^2-\Mp^2\Mpp^4+3\Mpp^6\over(\Mp^2-\Mpp^2)^3}
	\ln{\Mp^2\over\Mpp^2}\biggr]\ ,
\end{eqnarray}
\end{mathletters}
where $F$ is defined by
\begin{equation}
F(M_1^2,M_2^2)=M_1^2+M_2^2-2{M_1^2M_2^2\over M_1^2-M_2^2}
	\ln{M_1^2\over M_2^2}\ .
\end{equation}
Since $F(M_1^2,M_2^2)\ge0$ and vanishes only when the masses are
degenerate, we see that $T\ge0$ and parametrizes the size of custodial SU(2)
breaking.  $S$ vanishes when the dileptons are degenerate, but can pick up
either sign when the masses are split.  While $U$ does not play as
important a role in confronting experiment \cite{peskin}, we note that the
dilepton doublet gives $U\le0$.  This result is the opposite of that found
for a chiral fermion doublet where $U$ is non-negative.

A complete calculation of $S$ and $T$ must take into account the mixing
between the SU(2) singlet and doublet Higgs.  This is especially important in
light of the upper limit on the $\331$ breaking scale which puts a non-zero
lower bound on the mixing.  Because of the
mixing, the dileptons and physical Higgs combine in their contributions.
For $S$, we find the full result
\begin{eqnarray}
\label{eq:fullS}
S=-{1\over\pi}\biggl[&&{2\over3}\sspp-{1\over3}\ssp
	+{9\over4}\ln{\Mp^2\over\Mpp^2}\nonumber\\
&&+{1\over4}\sspp\ln{\mpp^2\over\Mpp^2}
	-{1\over4}\ssp\ln{\mp^2\over\Mp^2}\nonumber\\
&&-\sspp\ccpp G\biggl(\frac{\mpp^2}{\Mpp^2}\biggr)
       -\ssp\ccp G\biggl(\frac{\mp^2}{\Mp^2}\biggr)\biggr]\ .\qquad
\end{eqnarray}
The function $G$ is defined by
\begin{equation}
G(x)={7x^2-38x-29\over36(x-1)^2}
+{x^3-3x^2+21x+1\over12(x-1)^3}\ln{x}\ ,
\end{equation}
and vanishes when $x=1$.  $G$ is positive when the Higgs are heavier than the
dileptons and is usually negative when they are lighter.  We see that the
Higgs corrections always enter with a factor of either $\spp$ or $\sp$ and
arise because of the mixing of scalars with different hypercharges.  As a
result, $S$ reduces to Eqn.~(\ref{eq:limSTU}) in the limit when the Higgs do
not mix.

Turning to $T$, we find that it has the general form
\begin{eqnarray}
\label{eq:fullT}
T={3\over16\pi\sin^2\theta_WM_W^2}\biggl[
	&&F(\Mp^2,\Mpp^2)\nonumber\\
&&+\ssp\ccp F(\mp^2,\Mp^2)\nonumber\\
&&+\sspp\ccpp F(\mpp^2,\Mpp^2)\nonumber\\
&&-\ssp\ccpp[F(\mp^2,\Mpp^2)-F(\Mp^2,\Mpp^2)]\nonumber\\
&&-\ccp\sspp[F(\mpp^2,\Mp^2)-F(\Mpp^2,\Mp^2)]\nonumber\\
+&&{1\over3}\ssp\sspp[F(\mp^2,\mpp^2)-F(\Mp^2,\Mpp^2)]\nonumber\\
+&&{4\over3}\ssp(\ssp-\sspp)(\mp^2-\Mp^2)\nonumber\\
+&&{4\over3}\sspp(\sspp-\ssp)(\mpp^2-\Mpp^2)\biggr]\ .
\end{eqnarray}
In deriving this, we had to use the relation $\ccp\Mp^2=\ccpp\Mpp^2$ implied
by the definitions of $\tpp$ and $\tp$.  Again, the Higgs corrections
come in only through their small mixing into an SU(2) doublet.  We find that
$T$ is positive in most of parameter space and becomes large when the Higgs
or dilepton masses are split greatly, thus breaking custodial SU(2).
A similar calculation for $U$ is straightforward, but since experimental
constraints on $U$ are not as strong, we do not present it here.

The full expressions for $S$ and $T$ depend on four unknown parameters of
the new physics --- the two dilepton masses and the two new Higgs masses
(the VEV angles are determined completely from the dilepton masses).
In order to understand the general behavior of these radiative corrections,
we now turn to three interesting cases:
(a)~the dileptons are degenerate in mass, $\Mp=\Mpp$;
(b)~the dileptons are maximally split in mass, $\sspp=0$; and
(c)~the Higgs and dilepton masses are related by $\mp=\Mp$ and $\mpp=\Mpp$.

(a) $\Mp=\Mpp$.
In order to give identical masses to $Y^\pm$ and $Y^{\pm\pm}$, the VEVs, $v$
and $v'$ must be equal.  As a result, $\tan\beta=1$ and
$\sspp = \ssp = M^2_W / 2\Mpp^2 $.  From Eqn.~(\ref{eq:fullS}), we find
for $S$
\begin{equation}
\label{eq:Sdegen}
S={1\over2\pi}{M_W^2\over \Mpp^2}\left[
	-{1\over3}+{1\over4}\ln{\mp^2\over\mpp^2}
	+\ccpp\left(G\left({\mpp^2\over\Mpp^2}\right)
	+G\left({\mp^2\over\Mp^2}\right)\right)\right]\ .
\end{equation}
Note that even when {\it all} masses are degenerate, $S$ takes on a non-zero
result.  In this case, we see that the singlet--doublet mixing in the scalar
sector gives rise to a negative $S$ \cite{dugan,lavoura}.  For large Higgs mass
splittings, the second term in (\ref{eq:Sdegen}) dominates, and $S$ is
positive for $\mp\gg\mpp$ and negative for $\mp\ll\mpp$.  From the fit in
the previous section, (\ref{STlimit}), we see that $\mp\alt\mpp$ is favored.

For $T$, we find the simple result
\begin{equation}
T={1\over16\pi\sin^2\theta_WM_W^2}\sin^4\alpha_{++}F(\mp^2,\mpp^2)\ ,
\end{equation}
which gives the bounds
\begin{equation}
0\leq T \leq {1\over 64 \pi\sin^2\theta_W}{M_W^2\over \Mpp^2}
            {\max(\mpp^2,\mp^2)\over\Mpp^2}\ .
\end{equation}
The lower limit corresponds to Higgs mass degeneracy and the upper limit to
large mass splitting.
{}From Eqn.~(\ref{STlimit}), we obtain the upper bound for
the heavier Higgs, namely $\max(\mp,\mpp) \leq 7.0{\rm TeV}$, for $\Mpp
\leq 350 {\rm GeV}$.

(b) $\sspp=0$.
Due to the VEV structure, the mass splitting of the dileptons is restricted
by the condition $|\Mp^2-\Mpp^2|\le M_W^2$.  The limiting case
$\Mp^2=\Mpp^2 + M_W^2$ can be realized by $v \gg v'$ or $\sspp\to0$.
In this case, the
doubly charged Higgs, which is $\Delta'^{\pm\pm}$, is a pure SU(2) singlet
and is not involved in the oblique corrections.

The parameter T is then given by
\begin{equation}
T=\cases{\displaystyle
{1\over16\pi\sin^2\theta_W}{M_W^2\over\Mpp^2}{\mp^2\over\Mpp^2}\ ,&
for $\displaystyle{\mp^2\over\Mpp^2} \gg 1$ ,\cr
\displaystyle-{3\over16\pi\sin^2\theta_W}{M_W^2\over\Mpp^2}\ ,&
for $\displaystyle {\mp^2\over\Mpp^2} \ll 1$ .\cr}
\end{equation}
We see that T can be negative if $\mp^2 \ll \Mpp^2$.
However, it is
negligible unless the dileptons are extremely light.  On the other hand, $T$
is always positive for heavy Higgs, $\mp^2 \gg \Mpp^2$.

Although the Higgs
contributions are induced by the small mixing, namely $\ssp = M_W^2/\Mp^2$,
we obtain a stringent bound for the Higgs mass, $\mp \leq 3.5 {\rm TeV}$,
for $\Mpp \leq 350 {\rm GeV}$.
If we take the other limit $v' \gg v$, then $\ssp \to 0$.  By the same token,
we find $\mpp^2 \leq 3.5 {\rm TeV}$.  Combining this with the case for $v =
v'$ in part (a), we expect the new charged Higgs
to be lighter than a few TeV.  Using both limits and the restriction on
the Higgs mass, we also find $|S|\alt0.3$
provided all new particles are heaver than $M_W$.
%
%

(c) $\mp = \Mp$ and $\mpp = \Mpp$.
Both expressions for $S$ and $T$ simplify considerably when the Higgs masses
are equal to the dilepton doublet masses.
Since the symmetry breaking hierarchy ensures that the mass splittings for the
dileptons and the Higgs bosons are small, we find
\begin{mathletters}
\begin{eqnarray}
-{23\over12\pi}{M_W^2\over \Mpp^2}\le &S&\le
{19\over12\pi}{M_W^2\over\Mpp^2}\ ,\\
0 \leq &T& \leq {1\over 16\pi\sin^2\theta_W}{M_W^2\over \Mpp^2} \ .
\end{eqnarray}
\end{mathletters}
For $\Mpp \geq 250{\rm GeV}$, we obtain $-0.06\le S\le0.05$ and
$0 \leq T \leq 0.009$ as expected for a small mass splitting.

When the $\eta$ sextet is taken into account, it introduces 12 additional
physical Higgs fields.  In this case the mixing between scalars
in different SU(2) multiplets becomes more complicated.  Nevertheless,
our conclusions that $S$ can pick up corrections due to the mixing of scalars
with different hypercharge and that $T$ measures the mass splitting between
scalars still hold.  Without any fine tuning in the Higgs sector, we
expect all physical Higgs to be lighter than a few TeV.

\section{Conclusions}
To summarize, we have examined both tree level $Z$--$Z'$ mixing and
one-loop oblique effects induced by the new charged gauge bosons and Higgs
bosons in the $\331$ model.  The precision experiments constrain the mixing
angle to be in the range $-0.0006<\theta<0.0042$ and gives
$M_{Z_2}>490\rm GeV$.
Additional indirect lower bounds can be placed on the $Z_2$ mass from both
FCNC considerations and from the $Z'$--dilepton mass relation.  The latter
gives the strongest limit and, along with the upper bound on the $\331$
scale highly restricts the neutral gauge sector of the model, giving
$1.4<M_{Z_2}<2.2\rm TeV$.

Constraints on the new Higgs bosons are obtained from examination of
the one-loop radiative corrections using the parameters $S$ and $T$.
The parameter $T$ can be negative for very light charged Higgs and is
positive for heavy Higgs.  Hence we obtain an upper bound for the new
charged Higgs masses, namely $\mpp , \mp \leq \hbox{a few TeV}$.  The
Higgs sector places strong constraints on the mass splitting between
the singly and doubly charged members of the dilepton doublet.  Hence
no restrictions can be placed on the dilepton masses past that
coming from the Higgs structure.  Nevertheless, other experiments, in
particular polarized muon decay \cite{carlson}, strongly restrict the
dilepton spectrum.

We note that in this model, it is possible to obtain (small) negative
values of $S$ and $T$.  This result is quite general and occurs because of
scalar mixing.
In order to obtain a negative $T$, there has to be mixing between different
SU(2) multiplets (in this case singlets and doublets).  Mixing of states with
different hypercharge also allows negative $S$ for the case when all masses
are degenerate.  These observations have also been made in
Ref.~\cite{lavoura}.

As the precision electroweak parameters are measured to higher accuracy, we
can start placing more stringent bounds on the new physics predicted by
this $\331$ model.  When the top quark mass is determined, it will
remove much uncertainty in the standard model contributions to $S$ and $T$;
the parameters then become much more sensitive to truly new physics.
Because the masses of the new particles are already tightly
constrained, both direct and indirect experiments at future colliders
may soon realize or rule out this model.

\bigskip
\centerline{{\bf Acknowledgements}}
J.T.L. would like to thank Paul Frampton and Plamen Krastev for useful
discussions.  This work was supported in part by the U.S. Department of
Energy under Grant No.~DE-FG05-85ER-40219 and by the Natural Science and
Engineering Research Council of Canada.

\begin{table}
\caption{The experimentally measured values \cite{langacker2,pdb},
and standard model predictions \cite{peskin}
(for $m_t=150 {\rm GeV}$ and $m_H=1000 {\rm GeV}$)
used in the fit.}
\begin{tabular}{|c|c|c|} \hline
Quantity & experimental value & standard model \\ \hline
$M_Z$ (GeV) & $91.187 \pm 0.007$ & input   \\
$\Gamma _Z$ (GeV) & $2.491 \pm 0.007$ & $2.484$\\
$ R = \Gamma_{had}/ \Gamma_{l \bar{l}}$ & $20.87 \pm 0.07$ & $20.78$ \\
$\Gamma_{b \bar{b}}$ (MeV) & $373 \pm 9$ & $377.9$ \\
$A_{FB} (\mu)$ & $0.0152\pm 0.0027$ & $0.0126 $ \\
$A_{pol} (\tau)$ & $0.140\pm 0.018$ & $0.1297$ \\
$A_e(P_\tau)$ & $0.134\pm 0.030$ & $0.1297$ \\
$A_{FB} (b)$ & $0.093\pm 0.012$ & $0.0848$ \\
$A_{LR} $ & $0.100\pm 0.044$ & $0.1297$ \\
\hline
$M_W/M_Z$ & $0.8789 \pm 0.0030$ & $0.8787$ \\
$Q_W (Cs)$ & $-71.04 \pm 1.81$ & $-73.31 $ \\
$g^2_L$ & $0.2990 \pm 0.0042$ & $0.3001$ \\
$g^2_R$ & $0.0321 \pm 0.0034$ & $0.0302$ \\
\end{tabular}
\label{data}
\end{table}

\appendix*{}
In the electroweak sector, we can choose three independent precisely
measured parameters, $\alpha$, $G_F$ and $M_{Z_1}$, from which in
principle we can predict all the outcome of experiments in the $\331$
theory. Due to the presence of an additional gauge boson, $Z'$, and the
corresponding $Z$--$Z'$ mixing, the results of the standard model
predictions, which are written in terms of the starred
functions \cite{kennedy2},
need to be modified.  If we neglect the effects due to
any combinations of both the
$Z'$ parameters and the standard model radiative corrections, the results
can be expressed as shifts with respect to the starred functions.

For convenience, we can define a parameter, $s^2_0$, which is given by
\begin{equation}
s^2_0~(1~-~s^2_0)~=~\frac{\pi\ \alpha(M_{Z_1})}{\sqrt2\ G_F\
M^2_{Z_1}}\ .
\end{equation}
{}From the present data, $s^2_0~=~0.23146\pm0.00034$ is precisely known.
Because of the $Z$--$Z'$ mixing,  the mass of the $Z_1$ is shifted by a factor
\begin{equation}
\label{dmz}
\frac{\delta M_{Z_1}}{M_{Z_1}} = -\frac{1}{2}
               \frac{M^2_{Z_2}}{M^2_{Z_1}}\theta^2 \ .
\end{equation}
Hence, we obtain
\begin{equation}
\frac{M_W/M_{Z_1}}{M_{W_\ast}/M_{Z_\ast}}~=~1+\frac{1}{2}
	\frac{1-s^2_0}{1-2s^2_0}
        \frac{M_{Z_2}^2}{M_{Z_1}^2}\theta^2\ .
\end{equation}

(i) $Z$--pole physics.  The gauge interaction of the light neutral gauge
boson, $Z_1$, is given by
\begin{equation}
\label{z1}
{\cal L}~=~\frac{e_\ast}{c_\ast s_\ast} \sqrt{Z_Z(f)}
  ~{Z_1}_{\mu} \left[ J^{\mu}_3(f)~-~Q(f)s^2_{\eff}(f)~J^{\mu}_V
        \right] \ , \\
\end{equation}
\begin{eqnarray}
\label{seff}
\delta s^2_{\eff}(f)~&=&~s^2_{\eff}(f)~-~s^2_\ast \nonumber \\
             &=&~\left[\frac{a^f+b^f}{Q(f)}-\frac{2b^f}{T_3(f)}\right]\theta
       \ ,\\
\noalignand
\frac{\delta Z_Z(f)}{Z_{Z_\ast}}~&=&~\frac{4b^f}{T_3(f)}\theta+
  \frac{M^2_{Z_2}}{M^2_{Z_1}}\theta^2 \ ,
\end{eqnarray}
where $a^f$ and $b^f$, given in Ref.~\cite{ng}, are the vector
and axial vector coupling coefficients for the $Z'$.
Therefore, the partial
width for the $Z$--boson relative to the standard model prediction is
given by
\begin{equation}
\label{width}
\frac{\Gamma(Z \rightarrow f \bar f)}{\Gamma(Z \rightarrow f \bar f)_\ast}~=~
 1~+~\frac{\delta Z_Z(f)}{Z_{Z_\ast}}
  -2Q(f) \frac{g_{V}(f)_\ast}{g^2_{V}(f)_\ast+g^2_{A}(f)_\ast}
  \delta s^2_{eff}(f)\ ,
\end{equation}
where ${g_V(f)}_\ast = \frac{1}{2} T_3(f)~-~Q(f){s^2}_\ast$ and
${g_A(f)}_\ast = -\frac{1}{2}T_3(f)$.
For $\Gamma(Z \rightarrow b \bar b )_\ast$, we also include the vertex
correction due to the top-quark.

By the same token, we can express the polarization asymmetry of
fermion $f$ as
\begin{eqnarray}
\frac{A_{pol}(f)}{{A_{pol}(f)}_*}~&=&~1~- \delta A(f)\ ,\\
\noalign{\vbox{\vskip\abovedisplayskip \hbox{with} \vskip\belowdisplayskip}}
\delta A(f)~&=&~~\frac{Q(f)}{g_{V}(f)_\ast}
   \frac{g^2_{A}(f)_\ast-g^2_{V}(f)_\ast}{g^2_{A}(f)_\ast+g^2_{V}(f)_\ast}
     \delta s^2_{eff}(f)\ .
\end{eqnarray}
Hence we obtain
\begin{eqnarray}
\frac{A_{pol}(\tau)}{A_{pol}(\tau)_\ast}~&=&~1-\delta A(\it l) \\
\frac{A_{FB}(\mu)}{A_{FB}(\mu)_\ast}~&=&~1-2 \delta A(\it l) \\
\frac{A_{FB}(b)}{A_{FB}(b)_\ast}~&=&~1-\delta A(b)-\delta A(\it l)\ .
\end{eqnarray}

(ii) low energy experiments.  The low energy interaction Hamiltonian is
given by \cite{ng}
\begin{equation}
\frac{4G_F}{\sqrt{2}} \left( 1+\frac{M_{Z_2}^2}{M_{Z_1}^2}\theta^2
\right) \left[ J_\mu J^\mu~-~2 \theta J'_\mu J^\mu +
\frac{M_{Z_1}^2}{M_{Z_2}^2} J'_\mu J'^\mu \right]\ .
\end{equation}
Therefore the effective left- and right-handed coupling coefficients for
neutrino scattering are modified to be
\begin{equation}
\label{R}
\epsilon_\lambda(q)~=~g^0_\lambda(q)_\ast
  (1+\frac{M_{Z_2}^2}{M_{Z_1}^2}\theta^2-4\theta a^\nu )
    -(\theta-4a^\nu \frac{M_{Z_1}^2}{M_{Z_2}^2})(a^q+\eta(\lambda)b^q)\ ,\\
\end{equation}
where $\eta(\lambda) = 1$ and $-1$ for $\lambda = R$ and $L$
respectively.
Hence we obtain
\begin{eqnarray}
\frac{g^2_\lambda}{{g^2_\lambda}_\ast}
    =&&\frac{\epsilon_\lambda(u)^2+\epsilon_\lambda(d)^2}
    {g^0_\lambda(u)^2_\ast+g^0_\lambda(d)^2_\ast} \nonumber \\
=&& 1+ 2(\frac{M_{Z_2}^2}{M_{Z_1}^2}\theta^2-4\theta a^\nu)\nonumber \\
 && -2\frac{g^0_\lambda(u)_\ast(a^u+\eta(\lambda) b^u)
        +g^0_\lambda(d)_\ast(a^d+\eta(\lambda) b^d)}
    {g^0_\lambda(u)^2_\ast+g^0_\lambda(d)^2_\ast}
   (\theta-4a^\nu \frac{M_{Z_1}^2}{M_{Z_2}^2})\ ,
\end{eqnarray}

For atomic parity violation, the weak charge is given by
\begin{eqnarray}
\frac{Q_W}{{Q_W}_\ast}&=&1 + \frac{M_{Z_2}^2}{M_{Z_1}^2}\theta^2
   + \frac{\delta C_1(u)(2Z+N)+  \delta C_1(d)(Z+2N)}
     {g^0_A(e)_\ast g^0_V(u)_\ast (2Z+N)+g^0_A(e)_\ast g^0_V(d)_\ast (Z+2N)}
\ ,\\
\noalign{\vbox{\vskip\abovedisplayskip \hbox{where}
\vskip\belowdisplayskip}}
\label{dC1}
\delta C_1(q)&=&- \theta (g^0_A(e)_\ast a^q+b^e g^0_V(q)_\ast)
   + \frac{M_{Z_1}^2}{M_{Z_2}^2}b^eb^q\ .
\end{eqnarray}
The quantities $g^0_{{R,L}_\ast}$ and $g^0_{{V,A}_\ast}$
in Eqs.~(\ref{R})--(\ref{dC1}) are evaluated at zero energy.

\figure{90\% C.L.\ allowed region in $(M_{Z_2},\theta)$ parameter space.
The dotted lines indicate the constraints from the minimal Higgs structure.
Also included are the FCNC bound of Ref.~\cite{ng}, the lower bound from
the $Z'$--dilepton mass relation, and the upper bound on $\331$ unification.
\label{fig1}}

\figure{Best fit point (cross) and 90\% C.L.\ contour in the $S$--$T$ plane for
the $\331$ model (solid line).
For comparison, the model independent
(oblique parameters only) fit to the same data is also shown (dotted
line).  $S=T=0$ corresponds to the reference point $m_t=150 {\rm GeV}$
and $m_H=1000 {\rm GeV}$.  $U$ is always taken as a free parameter.
\label{fig2}}


\begin{references}
\bibitem{frampton} P. H. Frampton, Phys. Rev. Lett. {\bf 69}, 2889 (1992);
F. Pisano and V. Pleitez, Phys. Rev. {\bf D46}, 410 (1992).
\bibitem{ng} Daniel Ng, {\sl The electroweak theory of SU(3) $\times$ U(1)},
Triumf preprint TRI-PP-92-125 (December 1992).
\bibitem{massb} This upper bound comes from imposing the condition
$\alpha_X<2\pi$ at the $\331$ scale, assuming a three Higgs doublet model
below it and using the normalization of Ref.~\cite{ng}.  An absolute upper
limit on the unification scale comes from
$\sin^2\theta_W<1/4$, giving $M_{Z_2}<3.2\rm TeV$ with corresponding
$M_Y<590\rm GeV$.
\bibitem{langacker} Paul Langacker and Mingxing Luo, Phys. Rev. {\bf D45},
278 (1992).
\bibitem{lep} The LEP Collaborations, Phys. Lett. {\bf B176}, 247 (1992).
\bibitem{charm} CHARM II Collaboration, Phys. Lett. {\bf B232}, 539 (1989).
\bibitem{peskin} M. E. Peskin and T. Takeuchi, Phys. Rev. Lett. {\bf 65},
964 (1990); Phys. Rev. {\bf D46}, 381 (1992).
\bibitem{altarelli} G. Altarelli and R. Barbieri, Phys. Lett. {\bf B253},
161 (1990).
\bibitem{marciano} W. J. Marciano and J. L. Rosner, Phys. Rev. Lett.
{\bf 65}, 2963 (1990); E: {\bf 68}, 898 (1992).
\bibitem{kennedy} D. C. Kennedy and P. Langacker, Phys. Rev. Lett.
{\bf 65}, 2967 (1990); E: {\bf 66}, 395 (1991).
\bibitem{holdom} B. Holdom and J. Terning, Phys. Lett. {\bf B247},
88 (1990).
\bibitem{golden} M. Golden and L. Randall, Nucl. Phys. {\bf B361},
3 (1991).
\bibitem{kennedy2} D. Kennedy and B. W. Lynn, Nucl. Phys. {\bf B322}, 1
(1989).
\bibitem{langacker2} P. Langacker, {\sl Precision Tests of the Standard
Model}, University of Pennsylvania preprint UPR-0555T, {\tt hep-ph/9303304}
(March 1993).
\bibitem{pdb} Particle Data Group, Phys. Rev. {\bf D45}, III.64 (1992).
\bibitem{carlson} E. Carlson and P. H. Frampton, Phys Lett. {\bf B283}, 123
(1992).
\bibitem{beltrami} I. Beltrami {\it et al.}, Phys. Lett. {\bf B194}, 326
(1987).
\bibitem{Hunt} J. F. Gunion, H. E. Haber, G. Kane and S. Dawson, {\it The
Higgs Hunter's Guide}, Addison Wesley, 1990.
\bibitem{2higgs} W. Hollik, Zeit. Phys. {\bf C37}, 569 (1988).
\bibitem{2higgs2} C. D. Froggatt and R. G. Moorhouse, Phys. Rev.
{\bf D45}, 2471 (1992).
\bibitem{dugan} M. J. Dugan and L. Randall, Phys. Lett. {\bf B264}, 154 (1991).
\bibitem{lavoura} L. Lavoura and L.-F. Li, {\sl Mechanism for obtaining a
negative $T$ oblique parameter}, CMU preprint CMU-HEP93-02 (January 1993).
\end{references}
\end{document}